# Asymmetric Nano Graphene Model Applied To Graphite-like Material Room-temperature Ferromagnetism

Norio Ota[1] Narjes Gorjizadeh[2] and Yoshiyuki Kawazoe[2]

[1] Hitachi Maxell Ltd., Osaka, Japan    [2] Tohoku University, Sendai, Japan

## Abstract

Room temperature ferromagnetic materials composed only by light elements like carbon, hydrogen and/or nitrogen, so called carbon magnet, are very attractive for creating new material categories both in science and industry. Recently several experiments suggest ferromagnetic features at a room temperature, especially in graphite base materials. This paper reveals a mechanism of such ferromagnetic features by modeling nanometer size asymmetric graphene molecule by using both a semi-empirical molecular orbital method and a first principle density function theory. Asymmetrically dihydrogenated zigzag edge graphene molecule shows that high spin state is more stable in total molecular energy than low spin state. Proton ion irradiation play an important role to create such asymmetric features. Also, nitrogen contained graphite ferromagnetism is explained by a similar asymmetric molecule model.

## 1, Introduction

Carbon based room temperature ferromagnet , so called carbon magnet, was a long term target for both science and industry. Carbon molecules associated with light elements like hydrogen and/or nitrogen is very attractive in view of ultra light organic magnet. It is a key for future ecological magnetic material and new spintronics application. These five years or so, several groups reported room temperature ferromagnetism in modified graphite materials. P.Esquinazi et al [1] predicted ferromagnetic magnetization hysteresis loop in proton ion irradiated highly oriented pyrolytic graphite (HOPG). Magnetic moments with orders of $5 \times 10^{-6}$emu was observed at a room temperature having a coercive force of 100Oe.They emphasis negligible Fe /Co contamination. Another important result of this paper is that saturation magnetization is proportional to the root of ion irradiated areal density. Saitama university group directed by Prof. N.Hiratsuka opened impressive photograph in a web home page [2], synthesized powder is magnetically attracted by a permanent magnet. Estimated Curie temperature is very high up to 800K. Their formal english paper [3] opens starting material as triethylamine, following direct pyrolysis at 925 C. resulting almost 0.5 emu /g saturation magnetization and graphite like X ray diffraction pattern. Those experiments encourage us to open a new door to carbon magnets. However, mechanisms and guiding principles how to design those materials are not clear yet.

Also these five years, another amazing discovery is single sheet graphene firstly reported by K.Novoselov et al.[4]. Extraordinary electronic transport properties nominate graphene as a post silicon [5][6], also extraordinary mechanical strength as a post iron steel [7]. Here we like to emphasis extraordinary high melting temperature over 3000 K, which is a necessary condition for a room temperature ferromagnetism.

There are several theoretical predictions on single layer infinite length graphene ribbon. Already in 1996, M.Fujita et al [8] applied tight binding model of pai-electron network to unlimited length ribbon resulting ferrimagnetic spin configurations at zigzag edge carbons, suggesting ferrimagnetism with total magnetization zero. In 2003, K.Kusakabe and M.Maruyama[9][10] proposed ferrimagnetic graphene ribbon model with non-zero total magnetization which has dihydrogenated carbon at one side zigzag edge, whereas another side is monohydrogenated. In their infinite length graphene ribbon, calculated unit cell contains one dihydrogenated carbon for a density function theory(DFT) based analysis.

To clarify the ferromagnetic features, we need actual two dimensional molecule model. Ferromagnetism appears as a total balance between edge carbon magnetic states. It is very necessary to calculate two dimensional spin mapping and comparing energy difference. Model molecule should have at least three edge carbons. For example, in case of dihydrogenated three carbons (six hydrogen atoms in total) ,there are two cases, one is three spins parallel (Quartet state),another is one spin remain(Doublet state). If a quartet state energy is lower than doublet, we can believe a stable ferromagnetism.

## 2, Piled Graphene Model

Proton ion irradiation experiment [1] is very suggestive to make a proper molecule model. Fig.1 shows our piled graphene model. Each element of HOPG is thought as one leaf of graphene molecule compared with autumn leaf fall on the ground one by one. Proton rain drops on piled graphene. Watching to second leaf from ground, only left side edge carbons are attacked by protons, whereas right side one are not by a shadow effect of upper layers. Fig.2 is a calculation model as an asymmetric nano graphene molecule, which is $C_{48}H_{21}$ having dihydrogenated three carbons at left side zigzag edge.

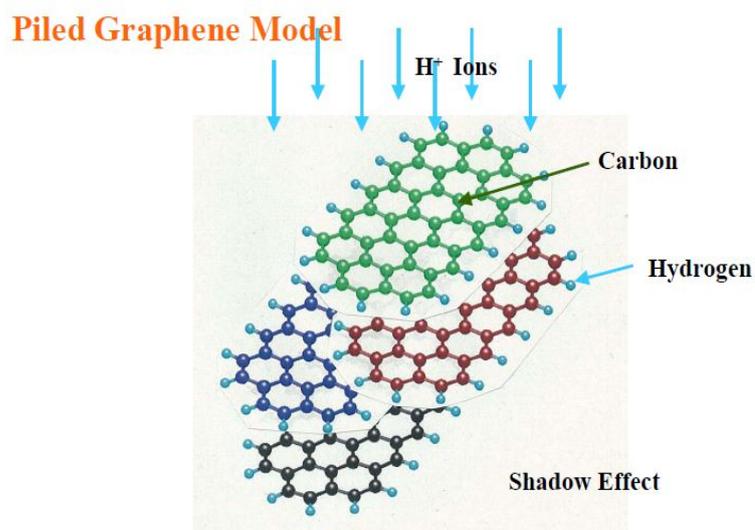

Fig.1 Graphene molecules are piled one by one., where proton ions are irradiated as like rain shower. Bare edge carbons are dihydrogenated, covered edge carbon remain as mono hydrogenated one.

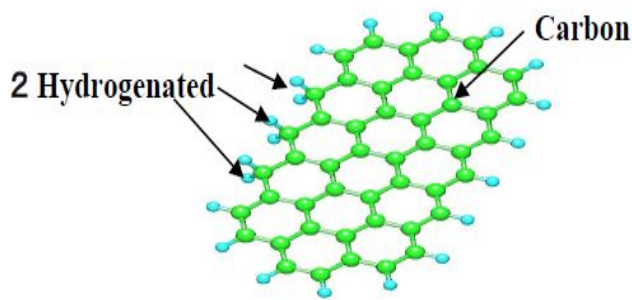

Fig.2 Asymmetric graphene molecule $C_{48}H_{24}$.. Left side zigzag edge three carbons are dihydrogenated, which is a model for calculation both on MO and DFT.

## 3, Calculation Methods

To clarify a magnetism of molecule, we have to obtain spin density mapping, square spin momentum= S(S+1) and total molecular formation energy to decide which magnetic state is stable. Two calculation methods are applied. One is a semi-empirical molecular orbital method(MO) : Fujitsu Scigress MO V1 Pro [11] employing PM5 function basis parameterized by many room temperature experiments. However, some consideration is necessary because of an insufficient electron-electron correlation in MO method so called spin contamination. Another is the first principle density function theory (DFT) based method: Gaussian DFT-UB3LYP [12] with a basis set of 6-31G, which give us fair electron-electron correlation, but limited on zero temperature ground state.

We need those two calculation results to estimate actual magnetic behavior at a room temperature.

## 4, Dihydrogenated Asymmetric Graphene molecule

### *4.1 Molecular Orbital Calculation*

It is well known that carbon atom has two bonding configurations, one is SP3 tetrahedral arrangement, another is SP2 arrangement. When dihydrogenation occurs at edge carbon ,SP3 configuration is favored remaining no spins on edge carbon. Whereas, mono-hydrogenation case, pai-electron remains as an isolated spin. Those basic difference affects to spin configuration and total behavior of molecular magnetism.

Fig.3 is a spin density mapping obtained by MO method. Up spins (dark gray: red in color) and Down spins (light gray: blue) are both peanuts like figures reflecting pai-orbitals and alternatively distribute each other. As illustrated in (a) , dihydrogenated three carbons(green) has no spins at all resulting total magnetization 3 $\mu B$ （Bohr Magneton）. In a computer, we can obtain Ms=1 $\mu B$ spin distribution as in (b). It should be noted that the difference of total molecule formation energy is,

$$E(3\mu B) - E(1\mu B) = -16.7 Kcal/mol$$

Such minus value implies that high spin state is more stable than lower one, in other words, larger magnetization will be realized in a molecule. We can understand why low spin state needs high energy by looking a spin distribution in (b). Inside of molecule, there is down-down spin pair generating higher energy.

MO method give us a simple understanding of magnetic behavior, but insufficient spin configuration. For example, S(S+1) of quartet state is overestimated as 5.5. Reasonable one is around 3.75 because of S value close to 1.5.

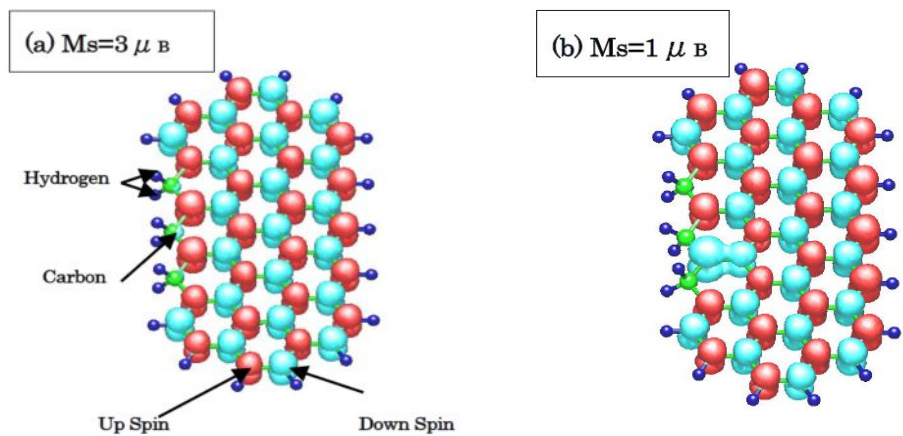

Fig.3  Mapping of Spin density contour surface obtainedby a semi-empirical molecular orbital method (scigress MO).Dark gray (red in color) shows up spin, while light gray(blue) down spin. Used function basis is PM5 parameterized by room temperature experiments. Figure (a) is Quartet state mapping, while (b) is doublet state. Contour surface  is 0.01 $\mu B/A^3$.

## 4.2 Electron-electron correlation

For a detailed magnetic behavior analysis, electron-electron correlation, mostly exchange interaction, should be calculated. Top figure in Fig.4 is an image of symmetric molecule spin configuration origin to pai-electron cloud(symbolic as butterfly mark) in MO level analysis.

When considering an electron-electron interaction, up spins will be exchanged with neighboring down spins resulting no net moment, especially in cyclic carbon molecule there appears diamagnetism as like aromatic molecules. Whereas in asymmetric case, there occurs some possibilities, one is ferrimagnetic configuration as predicted by M.Maruyama and K.Kusakabe [10] in unlimited length ribbon as imaged in (A), others are ferromagnetic configuration as shown in (B)and (C) depending on a distance between atoms and strength of exchange interaction.

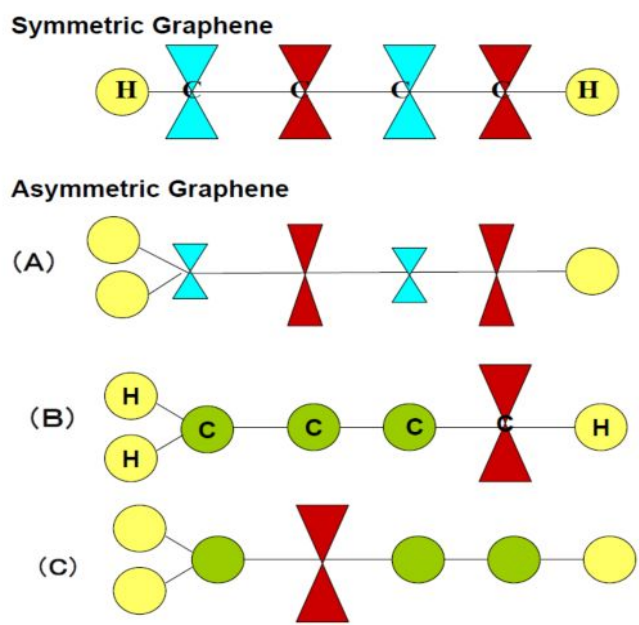

Fig.4　Cross sectional images of pai electron originate spin arrangement. Symmetric graphene in top view shows up spin(dark gray buttefly : red in colour) and down spin (light gray: blue)　in Hartree-Fock　level modeling. In an Asymmetric case, there are several magnetic capabilities when considering electron-electron correlations as like　Ferrimagnetic arrangement (A) and Ferromagnetic ones (B)and (C).

## 4.3 Density Function Theory Based Analysis

　　DFT based analysis is very helpful to decide which spin arrangement is favorable for a molecule. To understand a fundamental spin configuration, only one zigzag edge carbon is dihydrogenated as shown in Fig.5.
We can see a ferromagnetic Up-spin major distribution close to an image in Fig.4 (B).
　　It looks as if ferromagnetic, but we should take care
that exchange interaction depends on the electron density. Fig.5 is only a spin contour surface at　0.01　$\mu \mathbf{B}/\mathrm{A}^3$.
We should check less spin density contour map. Fig.6 is a result of 0.001 $\mu \mathbf{B}/\mathrm{A}^3$. Up and Down spins are distributed one by one as like ferrimagnetic one. However, we should notice that no spins at dihydrogenated carbon and right hand side spins grow larger affected by deeper ferromagnetic like spin density. Eventually, total map is a sum of ferromagnetic distribution and ferrimagnetic one.

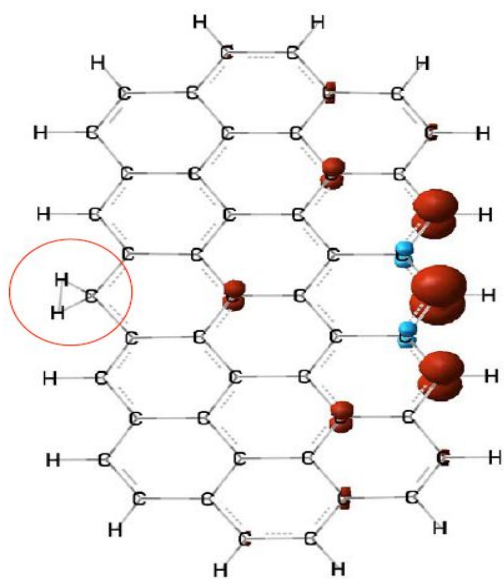

Fig.5　　For a simple understanding, only one carbon is dihydrogenated (circled one). Spin density map looks like as if Ferromagnetic in spin contour surface map at 0.01 μB/A³.

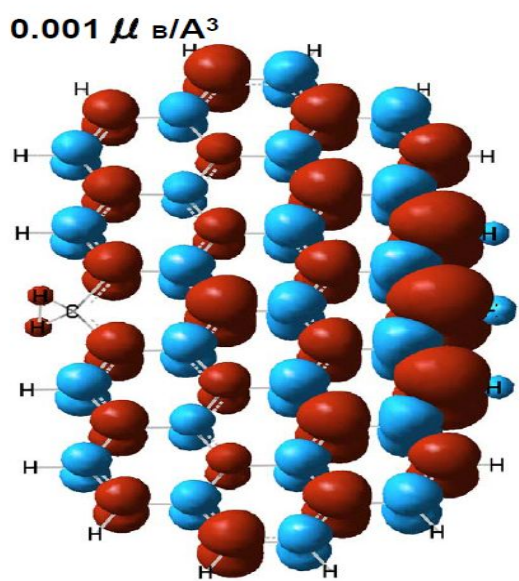

Fig.6  In case of contour surface at 0.001μB/A³,it looks as like the sum of Ferrimagnetic distribution  and Ferromagnetic one.

For a comparison with MO , also three carbons are dihydrogenated . By a DFT calculation, spin maps are obtained as shown in Fig.7(a) and (b) . We can see very complex spin arrangement in (b). Detailed calculated values are summarized in Table1.
It is important to compare total energy difference .
Result is,
$$E(3\mu B) - E(1\mu B) = -7.9 \text{ Kcal/mol}.$$
Again, larger magnetic moment configuration(Quartet state:3μB) is more stable than smaller one(Doublet state：1 μB) but almost half value comparing with MO result.. In DFT calculation, there appears remained spins on hydrogen atoms bonded to edge carbons.

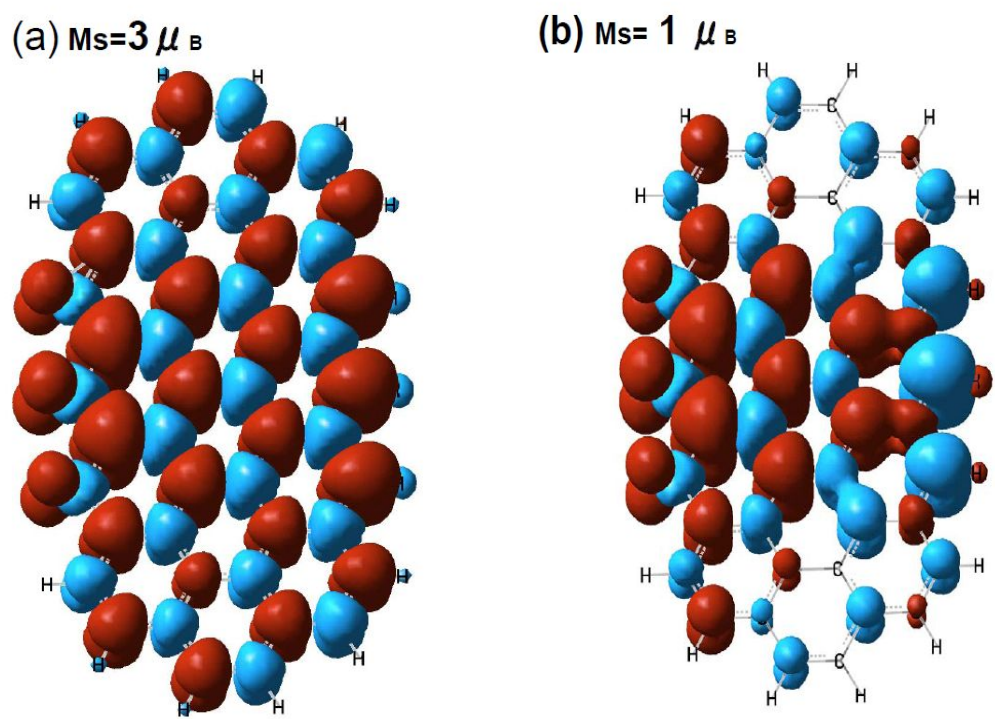

Fig.7  Spin density mapping of  Quartet state (a) and Doublet state(b) of $C_{48}H_{24}$ asymmetric nano graphene molecule calculated by a Gaussian

density function theory based UB3LYP using 6-31basis. Spin contour is 0.001 $\mu_B/A^3$.

Table 1, Calculated results of graphene molecule  —  Gaussian DFT UB3LYP, 6-31Gbasis

|  | $C_{48}H_{21}$ | $C_{48}H_{24}$ | | $C_{45}N_3H_{21}$ | |
| --- | --- | --- | --- | --- | --- |
| Spin State | Singlet | Quartet | Doublet | Quartet | Doublet |
| S(S+1) by DFT calculation | 0.0 | 3.76 | 0.96 | 3.75 | 0.75 |
| Energy Difference * | — | -7.90 kcal/mol | | +11.64 kcal/mol | |
| Molecule Size (width x length) | 0.928nm x 1.663nm | 0.924 x 1.679 | 0.924 x 1.679 | 0.917 x 1.665 | 0.913 x 1.663 |
| C-C, C-N bond length ** | 0.140nm | 0.149 | 0.149 | 0.141 | 0.139 |
| C-C-C,C-N-C bond angle ** | 122.0 degree | 115.3 | 115.3 | 118.9 | 122.0 |

* Energy difference = E (Quartet) – E (Doublet)
** Bond length, bond angle are at zigzag edge center atom

Table 1, Summary of major calculated results by a Gaussian DFT program using UB3LYP method with 6-31basis.

### 4.4 Proton Irradiation Dependence on Magnetization

Another important experimental evidence of Ref. (1) is that saturation magnetization Ms is proportional to the root of proton irradiation areal density D.

$$Ms = k\, D^{1/2} \quad \text{(from experiment)}$$

Our calculation suggests that Ms is proportional to the number ( n) of dihydrogenated carbon atoms on zigzag edge position.

By a shadow effect modeled in Fig.1, the number : n is proportional to a linear irradiation density : d, not an areal
density: D. Relationship d = $D^{1/2}$ is obvious, therefore

$$Ms = n\, \mu_B = k\, d = k\, D^{1/2} \quad \text{(from calculation)}$$

Thus, we can simply explain proton irradiation results.

### 4.5 Nitrogen substitution to edge carbon

Graphite based another experimental example was opened by K.Kamishima et al. (2).

Saturation magnetization is relatively large almost 0.5emu/g at a room temperature. We tried to explain again by an asymmetric graphene molecule model.

Their starting material is Triethylamine. After 975 C heating, X-ray analysis shows graphite like peak. Based on those information, we modeled again an asymmetric molecule $C_{45}N_3H_{21}$ as shown in Fig.8. Three carbon atoms at left side zigzag edge are substituted by three nitrogen atoms, which are all mono-hydrogenated respectively. Nitrogen has one lone pair electrons, up and down spin pair in a same orbit. There disappear local net moment. This is a similar situation with dihydrogenated carbon case. Therefore, we can expect asymmetric spin arrangement in a molecule. Fig.8 is a result for spin density contour at 0.01 $\mu B/A^3$ where (a) is for Ms=**3 $\mu$ B** and (b) for Ms=**1 $\mu$ B.**

It looks ferromagnetic like features whole on a molecule both on (a) and (b). Such features depend on excess electrons on nitrogen atoms simulated as in Fig.9, which shows very high density electrons (over $0.4e/A^3$) limited on nitrogen position, but less spin densities as shown in Fig.8(b). Those are basically considered as a lone pair electrons effect of nitrogen valence orbit.

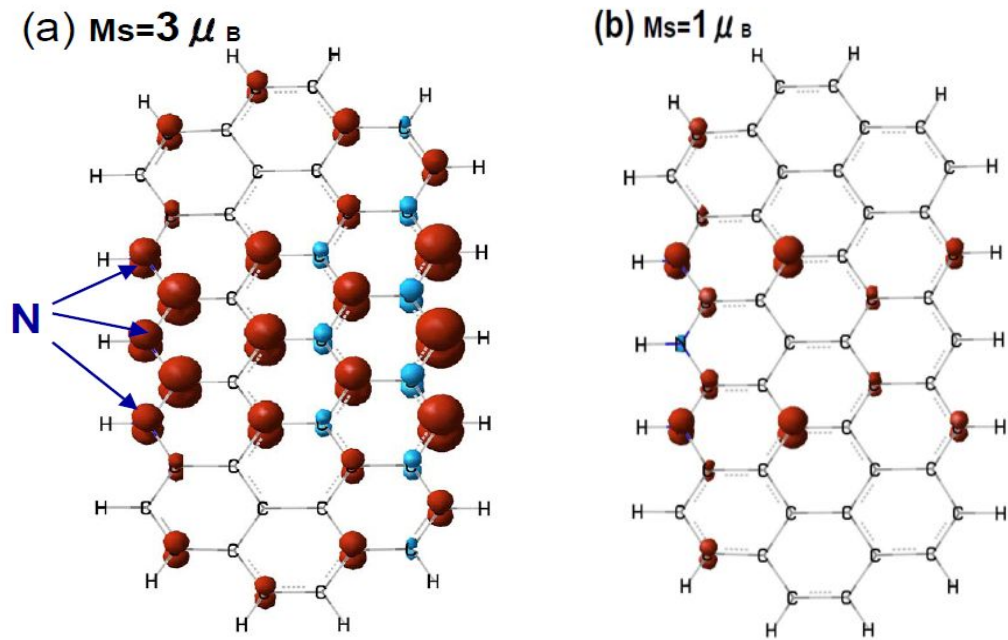

Fig.8  Three nitrogen atoms substitute left side zigzag edge carbons as **$C_{45}N_3H_{21}$**. Quartet(a) and Doublet(b) state are illustrated as spin contour surface at 0.01 $\mu B/A^3$.

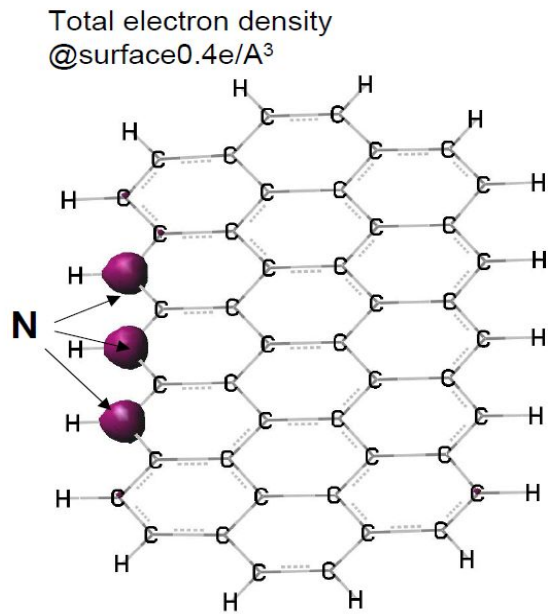

Fig.9  Total electron density map of **$C_{45}N_3H_{21}$** at a surface of 0.4e/A³. Excess electrons on nitrogen atoms cause an asymmetric spin configuration all over the molecule.

Energy stability between two magnetic states should be compared. Result is,
$$E(3\mu B) - E(1\mu B) = +11.6 \text{ Kcal/mol}$$
that is, low spin state (Ms=**1 $\mu$ B**) is more stable than high state one. Therefore, for obtaining larger magnetization in nitrogen substitution we should consider other capabilities as like zigzag edges substituted by C-N-C-N- chains .

## 5,  To Obtain Macro Scale Ferromagnetic Material

From experimental evidences and our calculation  results ,we can believe a possibility of room temperature ferromagnetism. However, our model is limited on a single molecule magnetic states. For designing a macro scale ferromagnetic materials, some self-organization mechanism combining so many molecules are necessary. If directions of each molecules are random, there only occurs superparamagnetic phenomena. Once we overcome such problems, asymmetric nano graphene has a potential to show around 50~100 emu/g magnetization assuming close packed molecules and magnetic anisotropy always perpendicular to all molecule plain. Those values are comparable to ferrite materials.

## 6, Conclusion

Carbon based material, especially in graphite modified system, has a potential to show a room temperature ferromagnetism. Asymmetric nano graphene molecule was modeled to clarify such mechanism and to obtain material design rule. Molecular orbital and density function theory calculation were applied to this model.

(1) Dihydrogenated zigzag edge molecule shows larger magnetization. Quartet state energy is lower and stable than doublet one.
(2) Proton irradiation dependence on magnetization is explained by a piled graphene model.
(3) We also tried nitrogen substitution to edge carbon resulting none or smaller magnetization behavior.

## 7, References